# Time-resolved dynamics of the superconducting two-gap state in $MgB_2$ thin films

Y. Xu,[1] M. Khafizov,[1(a)] L. Satrapinsky[1,2] P. Kúš,[3] A. Plecenik,[3] and Roman Sobolewski[1(b)]

[1]Department of Electrical and Computer Engineering and Laboratory for Laser Energetics, University of Rochester, Rochester, NY 14627-0231

[2]Institute of Electrical Engineering, Slovak Academy of Science, SK-84239 Bratislava, Slovak Republic

[3]Department of Solid State Physics FMFI, Comenius University, SK-84248 Bratislava, Slovak Republic

Femtosecond pump-probe studies show that carrier dynamics in $MgB_2$ films is governed by the sub-ps electron-phonon (e-ph) relaxation present at all temperatures, the few-ps e-ph process well pronounced below 70 K, and the sub-ns superconducting relaxation below $T_c$. The amplitude of the superconducting component versus temperature follows the superposition of the isotropic dirty gap and the 3-dimensional $\pi$ gap dependences, closing at two different $T_c$ values. The time constant of the few-ps relaxation exhibits a double-divergence at temperatures corresponding to the $T_c$'s of the two gaps.

PACS: 74.25.Gz, 74.70.Ad, 78.47.+p

---

(a) Also at the Department of Physics and Astronomy, University of Rochester, Rochester, NY 14627.
(b) Also at the Institute of Physics, Polish Academy of Sciences, PL-02668 Warszawa, Poland.

The discovery of superconductivity near 40 K in magnesium diborides[1] has stimulated very intensive investigations intended to fully understand the superconducting mechanism and the physical properties of these materials. The phonon-mediated, Bardeen-Cooper-Schrieffer (BCS) pairing mechanism was identified by the boron isotope effect,[2] as well as through other supporting experiments, including tunneling,[3-5] photoemission spectroscopy,[6,7] and nuclear magnetic resonance.[8] Recent theoretical work has predicted a two-band energy model for the $MgB_2$ superconductor, with two different BCS order parameters $\Delta$,[9] closing at the same critical temperature value $T_c$: the 2-dimensional $\sigma$ band gap $\Delta_\sigma$ with $2\Delta_\sigma/k_B T_c \approx 4$ and the 3-dimensional $\pi$ band gap $\Delta_\pi$ with $2\Delta_\pi/k_B T_c \approx 1.3$. It was stressed by Liu et al.[9] that these two different $\Delta$'s could be observed only in the clean limit, when the interband scattering is very weak. Indeed, the experimental data, obtained using the point-contact and scanning-tunneling spectroscopies, are in good agreement with the $\Delta_\sigma$ and $\Delta_\pi$ values predicted by the theory.[3-5] The anomalous behavior of the $MgB_2$ specific heat temperature dependence has also been explained within the two-gap model.[10]

In the dirty limit, due to strong interband and intraband scattering, one can typically measure an isotropic energy gap $\Delta_{dirty}$ with the zero temperature value of about 4 meV.[11] The temperature dependence of $\Delta_{dirty}$ deviates considerably from the BCS $\Delta(T)$ curve,[7,12] with a hump at temperatures below $T_c$.[13] Although the existence of the hump could be accounted for by introducing proximity-induced intergap coupling, the most natural explanation was presented using the two-gap model with two different $T_c$ values: $T_{c,dirty}$ corresponding to the BCS $\Delta_{dirty}(T)$ dependence and the specimen's $T_c$, related to closing of $\Delta_\pi$.[13] The detailed correlation between the $T_c$'s and the interband and intraband scattering mechanisms has been studied in the framework of the two-gap model by Mazin *et al,*[14] and this approach is most suitable for MgB$_2$ thin films fabricated using a post-annealing process, since they exhibit both the crystalline and amorphous features.

In this letter, we present time-resolved, pump-probe measurements of post-annealed MgB$_2$ thin films with very smooth surface morphology. The Mg-B precursor films were prepared by co-evaporation of Mg (purity 99.8%) and B (purity 99.9%) components on unheated *r*-cut sapphire and mica substrates. The deposited precursors were about 120 nm thick and were *ex-situ* annealed in a vacuum chamber, using a computer-controlled halogen-lamp heater.[15] The annealing temperature was increased at the rate of 40ºC/s for mica and 60ºC/s for sapphire substrates and was leveled at 720ºC for 300 s. Next, the halogen lamps were switched off and the chamber was filled with Ar to quench the samples to room temperature in 30 s. The resultant MgB$_2$ films were amorphous with nanocrystal inclusions[16] and exhibited optically smooth surfaces. For both mica and sapphire substrates, the onset of the superconducting transition $T_{c,\text{on}}$ was ~35 K and the transition width was approximately 3 K. The critical current density $j_c$ at 4.2 K was about $10^7$ Acm$^{-2}$.

Optical pump-probe measurements of transient reflectivity change $\Delta R/R$ were performed on MgB$_2$ thin films in the temperature $T$ range from 10 K to 300 K. The samples were mounted on a cold finger in a *T*-controlled, liquid-helium continuous-flow optical cryostat. The light source was a commercial Ti:Sapphire laser, which produced 100-fs-wide pulses at a repetition rate of 76 MHz. Since we did not observe any wavelength dependence of the $\Delta R/R$ signal within the tunable range (710 nm to 950 nm) of our laser, we chose 800-nm wavelength (1.5-eV photon energy) to be our working optical radiation. The pump and probe beams were cross-polarized to avoid the coherent artifact, with an energy ratio of at least 10:1. The spot sizes on the samples were <100 $\mu$m in diameter. The probe-beam reflected from the samples was collected by a photodetector and measured using a lock-in system. The pump energy was always smaller than 26 pJ per pulse (~2-mW average power) to avoid sample heating.

Figure 1 shows a set of three $MgB_2$ photoresponse $\Delta R/R$ signals, each typical for a different $T$ range. In all cases, the $\Delta R/R$ rise time is limited by the width of the excitation pulses, while the relaxation processes change quite significantly with $T$. At high $T$'s (300 K to approx. 70 K), the signal is dominated by a fast, single-exponential decay with a time constant $\tau_1 \approx 160$ fs, as the electron system loses its high excess energy through the electron-phonon (e-ph) interaction.[17] This behavior is in good agreement with previous pump-probe measurements, performed on various metallic thin films at room temperature.[18] The sub-ps relaxation is followed by a component with a decay time $\tau_2$ of the order of several picoseconds. As demonstrated in Fig. 1, the amplitude of this few-ps process is almost negligible at high temperatures. However, it becomes significant in the photoresponse below 70 K, so that the $\Delta R/R$ signal is bi-exponential, as a combination of $\tau_1$ and $\tau_2$ relaxations, with $\tau_2$ strongly dependent on $T$. Finally, below $T_c$, a third process appears, in addition to the sub-ps and few-ps components existing above $T_c$. This third process is characterized by a $T$-dependent negative amplitude and a relaxation time $\tau_3$ on the order of hundreds of picoseconds, as shown in the inset in Fig.

1. The decay time $\tau_3$ is independent of the sample temperature, but it varies with different substrate materials.

To better understand the complex nature of the experimentally observed $\Delta R/R$ responses, we decomposed the total signals, using the nonlinear least squares fitting method into three separate single-exponential processes with different time constants of $\tau_1$, $\tau_2$, and $\tau_3$. Figure 2(a) shows that the bi-exponential $\Delta R/R$ observed between $T_c$ and 70 K (solid squares) can be fitted as the sum of two processes: the sub-ps e-ph process (dotted line), the same as observed at high $T$'s (see top curve in Fig. 1) and is characterized by the same $\tau_1 \approx 160$ fs, and the few-ps e-ph process (dash-dotted line) with the $T$-dependent $\tau_2$. For the completeness of simulations, we also added a small, very slow offset (dashed line), which is a signature of not studied here, long-lasting phonon cooling process.[18] Figure 2(b) demonstrates that the experimental $\Delta R/R$ at temperature far below $T_c$ (solid squares) consists of the two e-ph relaxation processes (dotted and dashed-dotted lines) observed above $T_c$, and the third process with a negative amplitude and a sub-nanosecond time constant $\tau_3$. The solid lines shown in Figs. 2(a) and 2(b) are superpositions of the three individual, single-exponential relaxation fittings and reproduce the experimental data extremely well.

In this work, we focus on the relaxation processes characterized by the $\tau_2$ and $\tau_3$ time constants, as the fast ($\tau_1$) e-ph process represents the initial cooling of very hot electrons and, as we stressed before, its nature is well understood.[17,18] The slowest ($\tau_3$) process with the negative amplitude is clearly related to superconductivity in $MgB_2$ since it only exists below $T_c$ and represents the decrease in the total number of excited electrons as they annihilate, forming Cooper pairs. The time evolution of the negative $\Delta R/R$ component is $T$ independent and is characterized by $\tau_3 \approx 400$ ps for ~100-nm-thick $MgB_2$-on-sapphire films, and by a somewhat shorter $\tau_3$ value for the $MgB_2$-on-mica films. Thus, we can identify $\tau_3$ as the phonon escape time $\tau_{es}$.[19] We must stress that the negative $\Delta R/R$ component in the superconducting state was reported earlier in a lot of works devoted to pump-probe studies of high-temperature superconductors (HTS).[20,21] In our $MgB_2$ films, this superconducting component could be seen only under a very weak pump perturbation. With high incident pump power (e.g., 5 mW of average power), the $\Delta R/R$ signal below $T_c$ contained only the positive components[22] [analogous to Fig. 2(a)], as the optical energy transferred from the electron system to phonons was large enough to drive the phonon temperature above $T_c$ during the early relaxation stage. The negative component was also absent at $T < T_c$ when our $MgB_2$ films were driven into their resistive state by a dc bias.[22]

The $T$ dependence of the amplitude of the negative $\Delta R/R$ component (closed squares) is shown in Fig. 3. The most striking is a hump-like behavior just below $T_c$.[13,23] As the negative component reflects the superconducting state of the material, one can expect that it should follow the $\Delta(T)$ dependence. The dashed line in Fig. 3 represents the BCS $\Delta(T)$ best fit. We note that at higher temperatures, our experimental data points deviate quite significantly from the BCS dependence. Fits based on the strong coupling formalism and the formula $\Delta(T) = \Delta(0)[1 - (T/T_c)^p]^{1/2}$ derived by Choi et al.[24] also failed to fit the experimental data.

The alternative approach is to fit the data with a combination of two BCS $\Delta(T)$ dependences (solid lines): one representing the isotropic $\Delta_{\text{dirty}}(T)$, closing at $T_{c,\text{dirty}} = 30.5$ K, and the other corresponding to the 3-dimensional $\Delta_\pi(T)$, closing at $T_c = 34.7$ K, equal to $T_{c,\text{on}}$ earlier obtained from the resistance versus temperature measurement. The above model should be applicable for the mixed clean/dirty case and is consistent with the fabrication procedure of our MgB$_2$ films.[13,14,23] Our films are definitively not in the clean regime, as is reflected by their relatively high normal-state resistivity [$\rho_{\text{DC}}(T = 40$ K$) > 10$ $\mu\Omega$cm],[14] and there must exist an enhanced intraband $\pi$ scattering. The "dirtiness" of our films result in appearance of the isotropic $\Delta_{\text{dirty}}$. On the other hand, $T_{c,\text{dirty}}$ is significantly larger than the 22-K value, predicted for the 100% dirty case.[9] Thus, we also observe a contribution from the clean state, through the presence of $\Delta_\pi(T)$, closing at $T_{c,\text{on}}$.

The picosecond ($\tau_2$) e-ph relaxation process exists both below and above $T_c$, and its time constant $\tau_2$ is strongly dependent on $T$ below 70 K, as shown in Fig. 4. First, we notice a divergence of $\tau_2$. The divergence of the relaxation time constant at $T_c$ was reported previously in different HTS materials.[25,26] It indicates an opening of the superconducting gap and is associated with the quasiparticle inelastic scattering and Cooper-pair formation, as the excited carriers pile up at the edge of $\Delta$. In our case, $\tau_2$ clearly diverges at $T_{c,\text{on}}$, but there is an anomaly in the $\tau_2(T)$ behavior that indicates a second, lower temperature divergence at $T$ corresponding to $T_{c,\text{dirty}}$. To confirm our findings, the inset in Fig. 4 shows the same $\tau_2(T)$ dependence measured in a different experimental run with more accurate temperature resolution–the double divergence is clearly visible. At the same time, no hysteresis in $\tau_2(T)$ with respect to ramping the temperature up or down was observed. Since the two temperatures at which $\tau_2$ diverges in Fig. 4 are the same as the $T_c$ values for the $\Delta_{\text{dirty}}$ and $\Delta_\pi$ gaps in Fig. 3, the double divergence in $\tau_2(T)$ directly supports our earlier conclusion that in our post-annealed $MgB_2$ films, the superconducting state is characterized by two superconducting gaps closing at two different $T_c$'s, with $\Delta_\pi$ being the main gap corresponding to $T_{c,\text{on}}$.

Above $T_c$, $\tau_2$ decreases gradually over a wide $T$ range from $T_c$ to 70 K. Our fitting procedure returns nonzero $\tau_2$ values even at temperatures above 90 K. But as we discussed in Fig. 1, the amplitude of this $\tau_2$ relaxation process above 70 K becomes very small and the dynamics of the $\Delta R/R$ signal in the >70-K range is dominated by the sub-ps e-ph relaxation. However, the fact that this few-ps relaxation component with the $T$-dependent $\tau_2$ extends to so far above $T_{c,\text{on}}$ is in direct contradiction with previous reports on HTS materials, in which $\tau_2$ would vanish within a few kelvins above $T_c$.[25,26] We have no solid explanation of this effect and tentatively interpret this few-ps relaxation process as an additional e-ph interaction. The e-ph coupling in $MgB_2$ is complicated, and the phonon modes are highly inharmonic.[9] Thus, the $\tau_1$ and $\tau_2$ relaxation processes could be associated as electron interactions with different phonon modes.

In conclusion, we have found that at room temperature and down to approximately 70 K, $MgB_2$ thin films respond to femtosecond optical photoexcitation as ordinary metallic thin films with a dominant sub-ps e-ph relaxation process. Below 70 K, a few-ps-long relaxation component becomes significant, indicating additional e-ph interaction. Finally, below $T_c$, there is a third photoresponse component, which we associate with the dynamics of the superconducting state and the recombination of quasiparticles into Cooper pairs. The time evolution of this latter component is sub-ns and corresponds to the phonon escape from the $MgB_2$ films. The temperature dependence of the amplitude of this superconducting component follows the $\Delta(T)$ anomalous temperature dependence, which can be interpreted for our $MgB_2$ films in the mixed clean/dirty limit as the two energy gaps, $\Delta_{dirty}$ and $\Delta_\pi$ closing at two different values of $T_c$, ($T_{c,dirty}$ and $T_{c,on}$, respectively). The same two $T_c$'s are also visible in the double-divergence of the relaxation time constant of the few-ps e-ph photoresponse component. Our work confirms that the 3-dimensional, small $\pi$ band gap is the intrinsic, bulk property of $MgB_2$ and can be observed even in the dirty case.

This work was supported by the US NSF grant DMR-0073366 and the NATO Linkage grant PST.CLG.978718 (Rochester), and by the Slovak Grant Agency for Science grants 2/7072/2000 and VEGA-1/9177/02 (Bratislava). Y. X. acknowledges support from the Frank Horton Graduate Fellowship Program in Laser Energetics.

**Figures**:

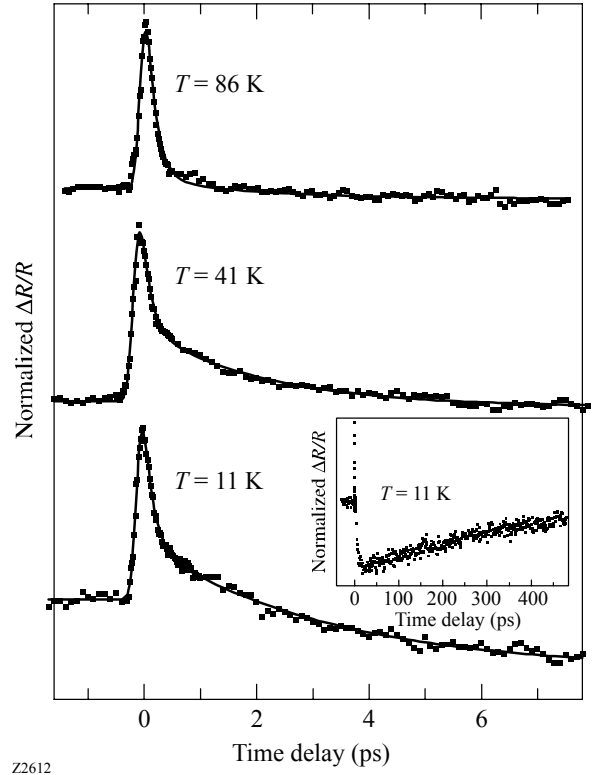

Fig. 1. Time resolved transient differential reflectivity $\Delta R/R$ of an MgB$_2$ thin film measured at three different temperatures. Both the pump and probe wavelengths were 800 nm and the average pump power was 1 mW. The inset shows the $\Delta R/R$ response below $T_c$ in a much longer time window.

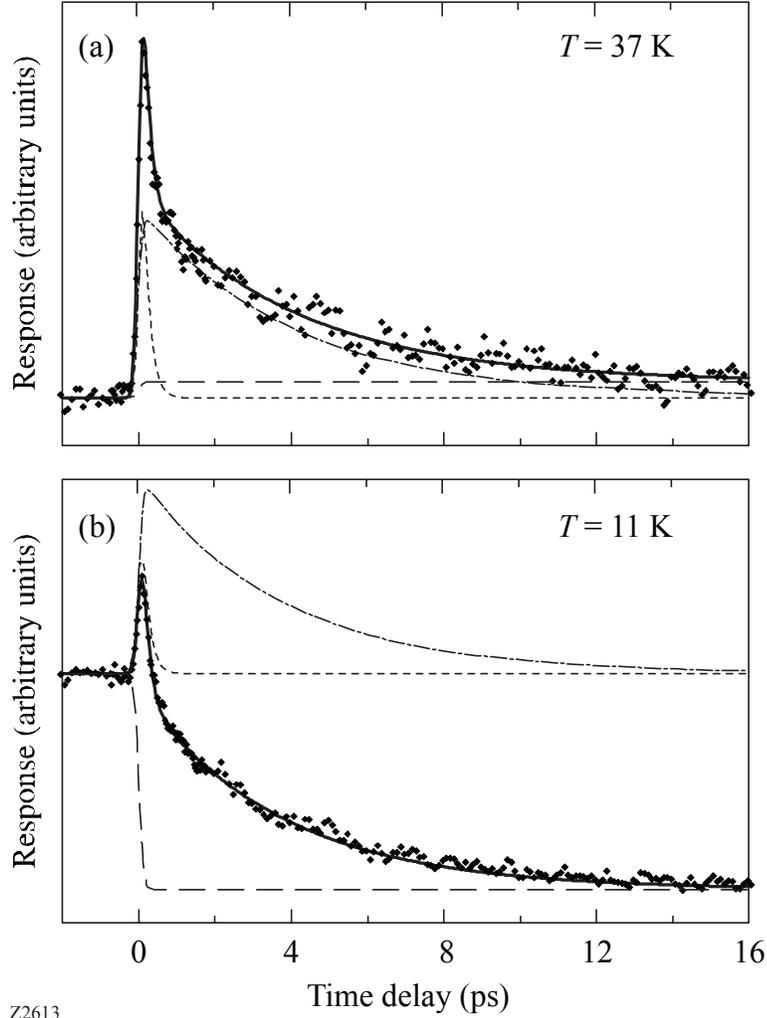

Fig. 2. The Δ*R/R* signals (solid squares) measured above (a) and below (b) $T_c$ and decomposed into constituting single-exponential relaxation processes. The dotted line represents the sub-ps e-ph process; the dashed-dotted line for the few-ps e-ph process; and the dashed line for the phonon relaxation either in the form of bolometric cooling (a), or the phonon escape from the superconducting film (b). Note that since $\tau_3$ of the phonon process is much longer than the time window shown, it appears to be a step function. The solid lines are the superpositions of the single-exponential processes.

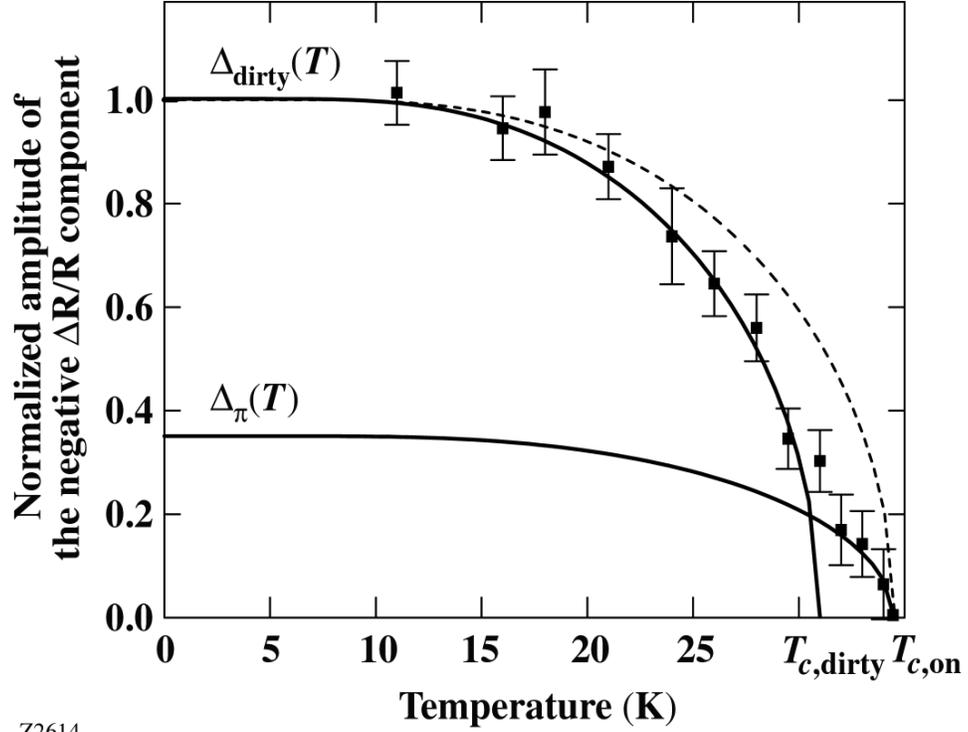

Z2614

Fig. 3. Temperature dependence of the amplitude of the negative ΔR/R component (closed squares). The dashed line is the BCS Δ(*T*) dependence; the solid lines are the two-gap model fits, each based on BCS theory. $T_{c,\text{on}}$ = 34.7 K and $T_{c,\text{dirty}}$ = 30.5 K. The error bars are error margins obtained from the least squares fitting procedure.

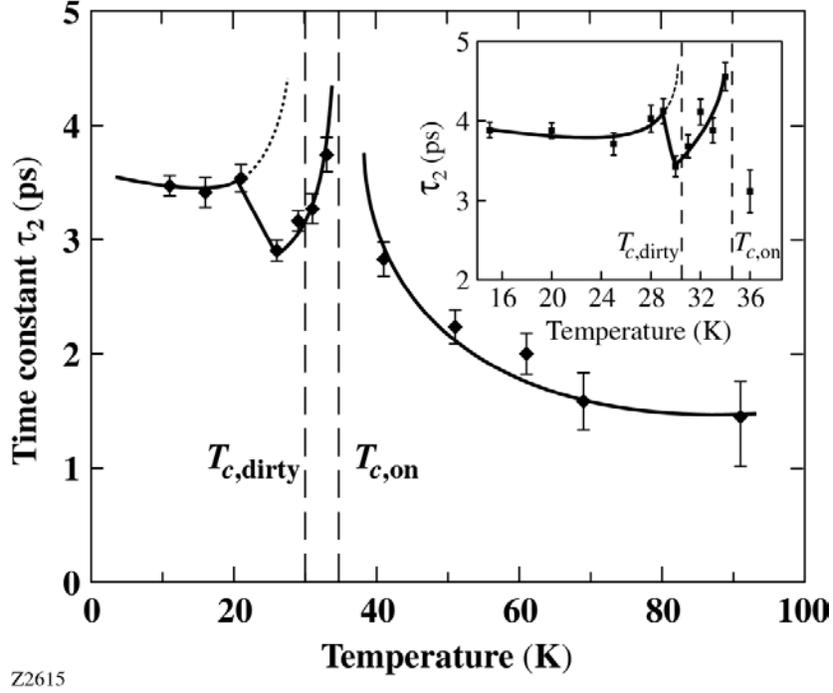

Fig. 4. Temperature dependence of the ∆R/R picosecond e-ph time constant $\tau_2$. The lines (solid and dotted) are only guides to the eye. The vertical dashed lines correspond to the $T_{c,\text{dirty}}$ and $T_{c,\text{on}}$ values obtained from Fig. 3. The inset shows the same double-divergence behavior observed in a different experimental run.